# Informal physics programs as communities of practice: How can program structures support university students' identities?

Brean Prefontaine[1], Claire Mullen[2], Jonna Jasmin Güven[3], Caleb Rispler[1], Callie Rethman[4], Shane D. Bergin[2], Kathleen Hinko[1], Claudia Fracchiolla[2]

[1] Michigan State University, East Lansing, Michigan, USA
[2] University College Dublin, Dublin, Ireland
[3]The University of Manchester, Manchester, UK
[4]Texas A&M, College Station, Texas, USA

Many undergraduate and graduate physics students participate in some form of public engagement throughout the course of their studies, often through groups supported by physics departments and universities. These informal teaching and learning programs can offer unique opportunities for physics identity development. Understanding how physics identities can be fostered will allow us to work toward a field that is inclusive of more identities. In this study, we build on previous work to investigate student-facilitator experiences in three informal physics programs using an operationalized Communities of Practice framework. Through our analysis, we identify different structures within these programs that support physics identity development.

## INTRODUCTION

Informal physics programs, activities, and events are learning experiences that happen outside of the formal classroom. Informal physics programs come in a variety of forms, are made for different audiences, and often have a variety of goals. Additionally, informal learning experiences offer agency to the learner and are free of many of the constraints seen in classrooms, such as standards and grades [1]. Many researchers and practitioners often refer to these informal physics environments as "outreach" or "public engagement" opportunities [2]. In this paper we use the term "informal" as a way to capture all of these different ways of engaging with physics outside of a formal learning environment.

The dynamism and diversity of informal spaces make them rich opportunities for physics education research. Informal learning environments are important to study because these physics spaces and programs can provide support to students of all ages in many different ways. Often, the research community has focused on the impact that participating in these programs has on the audience. However, we postulate that participation in these spaces also has an impact on those who facilitate them, especially undergraduate and graduate physics students [3-7]. We hypothesize that informal physics programs offer a space for positive physics identity formation in a way that differs from formal physics education, and thus these spaces can contribute to a more equitable field. Therefore, understanding how structures in informal physics programs impacts university students' physics identity is important for designing supportive learning environments in both informal and formal spaces.

In past work, we have found a variety of opportunities for growth and support are available to students who participate in facilitating informal physics programs, such as gaining teaching experience, developing communication skills, increasing content knowledge, and being able to feel like experts within their field of study [7-11]. Additionally, these programs also provide space for student facilitators to explore and develop physics identities [12]. By focusing on how physics identities are supported and fostered within these environments, we can better understand how to support students within their physics studies; additionally, cultivating a physics identity has been linked to persistence within the field [13-16]. An understanding of physics identity is particularly important for students who are historically marginalized and underrepresented in the field of physics because it can lead to a change in who sees themselves as a physicist and their sense of belonging within the field [17-20]. In part, this disengagement or lack of sense of belonging can be connected with tensions in students intersecting identities [18,19, 21-26] Previous studies have looked at what spaces/environments and structures allow students to bring forward their intersecting identities, including their physics identities, in both formal and informal environments [12,27-28].

The Community of Practice framework looks at identity formation from the socio-cultural perspective, therefore serving as a lens to understand the development and support of intersecting identities. Previous studies have used the Community of Practice framework to understand development of physics identity in formal learning spaces. In physics, Close et al. looked at how participation in the Learning Assistance program (LA) supported university students' physics identity [29,30]. They combined the mechanisms of identity from the Community of Practice framework with Hazari's Physics Identity constructs to determine what factors from the LA program impacted the students' physics identity [24]. A study by Irving and Sayre used the community dimensions of the Community of Practice framework to determine if participation in upper division physics labs impacted university physics students' level of membership in physics, i.e. shifts in their perceived identity development [31].

In prior work, we have postulated that informal physics programs can support physics identity by functioning as a community of practice for undergraduate and graduate student facilitators [3-4,6,12]. We operationalized Lave and Wenger's theoretical Communities of Practice framework so that it was contextualized to the informal physics environment [12]. We then utilized a case study approach to understand the individual experiences of university students who acted as program facilitators within an informal physics program. In using the Communities of Practice framework to understand the experiences of the university students, we were able to learn how individuals navigate their membership and the development of their physics identity within these communities.

In this paper we build on our prior work, which was focused on individual experiences, to consider the structures and norms of informal physics programs and how they contribute to experiences within communities of practice. We seek to answer the question, *"What are the structures and practices of informal physics programs that support physics identity development in physics student facilitators?"* To answer this question, we have conducted an in-depth investigation into three different informal physics programs that function as communities of practice and how they contribute to fostering university students' physics identities [32]. We looked at the programs through the lens of individual facilitators' experiences by collecting interviews with student facilitators. We have analyzed these interviews with our operationalized Communities of Practice framework for informal physics programs to understand how these programs may support identity development. Results from this study are relevant to both practitioners and researchers who are looking to improve the support given to university physics students and when creating new informal learning environments.

# COMMUNITIES OF PRACTICE FRAMEWORK

In 1991, Lave and Wenger developed the Communities of Practice framework to identify and study groups who come together with a shared concern or goal and who spent time together learning how to achieve that mission [33-34]. This theoretical framework is a social theory of learning and is useful for understanding how members within the group, and the group as a whole, learn and evolve to achieve a shared goal. The group develops a set of norms and practices that members of the group can recognize, practice, and influence in order to become part of the collective. In this sense, the Communities of Practice framework can be useful for understanding how a group operates in two different ways - how individuals experience the group and how the group functions as a whole. Furthermore, Lave and Wenger used this Communities of Practice framework to understand how participation impacts identity through membership among participants.

Not every group is considered a community of practice; rather, a group must contain three aspects to be considered a community of practice, as shown in Figure 1. The group must have some sort of shared goal or proficiency (the *domain*), interactions between groups members that help individuals achieve the goals (the *community*), and a set of common norms, repertoire, and shared information needed to achieve the goals (the *practices*) [34]. If a domain, community, and practice can be identified within a group of individuals, then it becomes appropriate to consider the group as a community of practice and to use the framework as an analysis tool.

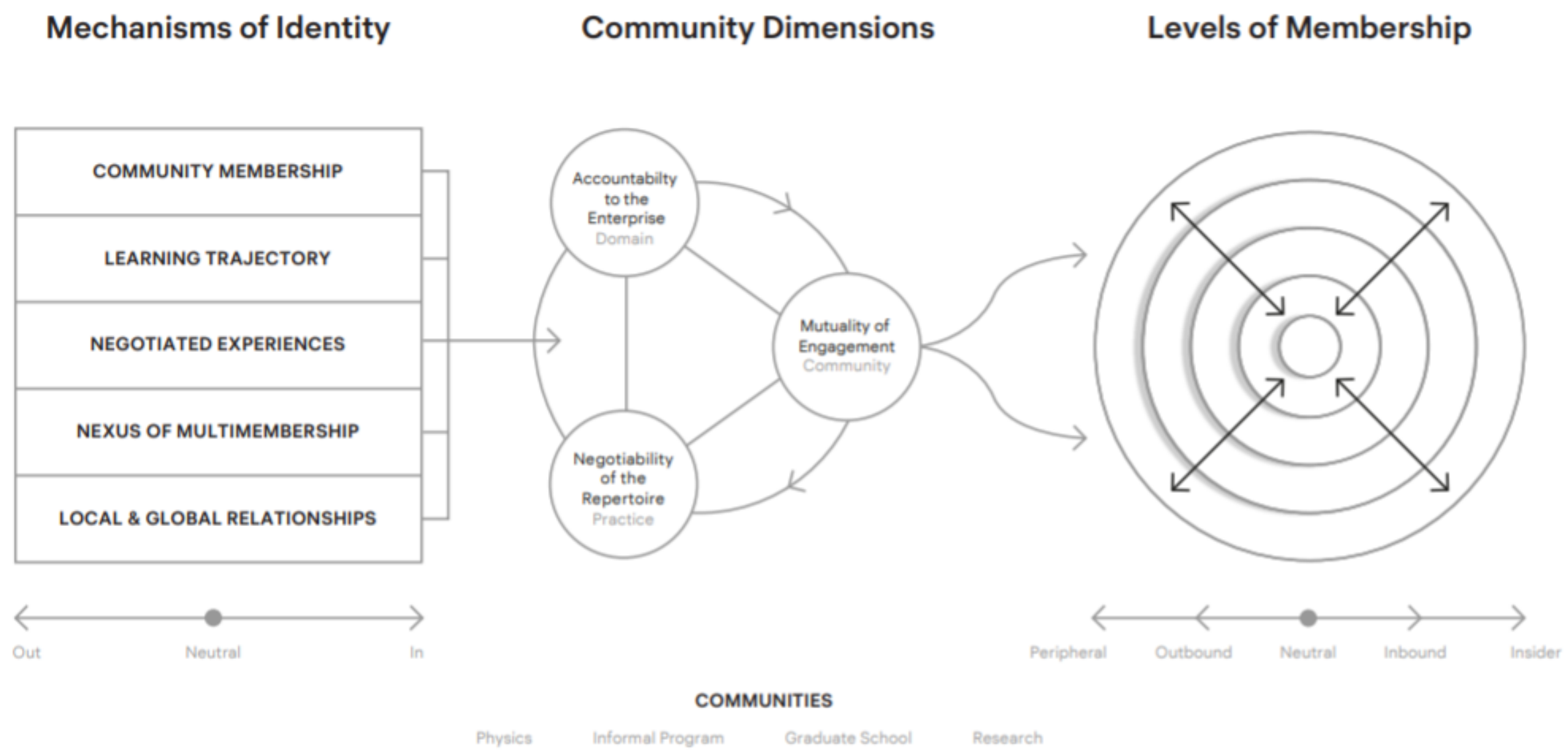

**Figure 1:** The three essential elements of a community of practice (shown in the middle column) contain the domain, community, and practice and must all be identifiable within a community in order to use the Communities of Practice framework. The mechanisms of identity (first column) interact with the elements of the community of practice and influence individuals' membership levels (shown in the third column).

If a group can be considered a community of practice, then it makes sense to consider both the roles individuals take within the community and also how individuals' identities are shaped and shape the community. In prior work, we operationalized these two aspects of Lave and Wenger theory in the context of informal physics programs. Our operationalized Communities of Practice framework [12] defines i) *Community Dimensions,* which help identify how the three aspects of a community of practice are recognized and impact an individual's membership in the community, and ii) *Mechanisms of Identity,* which identify the specific structures of informal physics programs that promote changes in an individuals' membership (i.e. if they move from being a peripheral member of the community to being a more central member). There are three community dimensions derived from what constitutes a community of practice: *Accountability to the Enterprise* is the commitment of individuals to the goals (domain) of the community. *Mutuality of Engagement* is the kind of interactions that an individual has with other members (the community), and *Negotiability of the Repertoire* is how an individual understands and engages in the activities of the group (the practices) [33-34]. These dimensions allow us to determine the role individuals perceive they have within the community of practice. A more detailed description of the *Community Dimensions* can be seen in Table 1.

Additionally, the framework can be used to understand what exactly is influencing changes in membership through the *Mechanisms of Identity* constructs. These constructs are: *Negotiated Experiences* are the ways individuals make meaning of experiences in the community, *Community Membership* is an individual's development of forms of competence and performance valued by the community, such as norms of interactions and the use of shared practices and resources, *Learning Trajectory* marks events that have taken place in the past or things that have been learned that resulted in different forms of participation within the community, *Nexus of Multimembership* highlights how members coexist and participate in different communities, and *Relationship between Local & Global* is an individual's negotiation of the sense of belonging within the local community and how that fits into a broader sense of more universal community [33-34]. Table 2 provides the theoretical definition of each of these constructs.

**Table 1:** The three *Community Dimension* constructs with definitions from the Communities of Practice framework.

| Community Dimensions | Definition |
|---|---|
| **Accountability to the Enterprise** | An individual's commitment and understanding of the goals and mission of the community. |
| **Mutuality of Engagement** | Interactions an individual has with other members of the community, such as peers, audience members, or directors/coordinators; also highlights how the individual is recognized or influenced by other community members. |
| **Negotiability of the Repertoire** | An individual's understanding of the practices of the community, including specific activities that the community performs to reach their goals and knowledge that the members need to have in order to be successful. |

**Table 2:** The five *Mechanisms of Identity* constructs with definitions from the Communities of Practice framework.

| Mechanisms of Identity | Definition |
|---|---|
| **Negotiated Experience** | The process of an individual making meaning of experiences through participation in the community and interactions with other members of the community. |
| **Learning Trajectory** | Events that have taken place in the past or things that have been learned that resulted in different forms of participation or changes in membership within the community. |
| **Nexus of Multimembership** | The participation and interactions of individuals with multiple communities, including the targeted community of practice, which requires coordination and negotiation. This construct highlights all forms of participation that contribute to our complete mesh of identities and how members coexist in different forms of membership. |
| **Community Membership** | The development of forms of competence and performance valued by the community, such as norms of interactions and the use of shared practices and resources. This mechanism identifies how more central members of the community become perceived as competent by other members and personally feel competent. |
| **Relationship between Local and Global** | The negotiation of the sense of belonging within the local communities and how that fits into a broader sense of more universal community. |

By using the *Community Dimensions* and *Mechanisms of Identity*, our operationalized Communities of Practice framework allows us to understand how membership of university students evolves throughout their facilitation of informal physics programs. Through participation, students may become more central members of the community or their level of commitment may lessen so that they become a more peripheral member. To identify membership levels, we turned to how Lave and Wenger discuss membership. The original Communities of Practice framework identified five membership levels - central, insider, neutral, peripheral, and outsider [34]. Parting from this element of their theory in order to make the coding more consistent, we created a set of subcodes during the operationalization process that identifies membership levels for each *Community Dimension* and the type of changes *Mechanisms of Identity* promoted. The *Community Dimension* constructs have five subcodes identifying membership levels - insider, inbound, neutral, outbound, and peripheral. Similarly, the *Mechanisms of Identity* have three subcodes that identify movement between membership levels - inbound, neutral, and outbound. For example, someone who has enjoyed participating and expresses the desire to continue participating in a certain community would experience identity development toward more central membership within the community (inbound), while someone who is not as deeply involved in the core activities may not experience this shift in identity (neutral or peripheral). A more detailed discussion of the Communities of Practice framework, the operationalized framework, and our methods of validation can be found in our previous work [12].

## ESTABLISHING INFORMAL PHYSICS PROGRAMS AS COMMUNITIES OF PRACTICE

In order to consider how informal physics groups may provide opportunities for identity growth, it is a necessary step to determine if an informal physics program does in fact have the characteristics of a community of practice before applying the framework to analyze group members' experiences. Informal physics activities can range widely in terms of format, frequency, content, audience, and level of involvement of university physics students; thus, it is likely that not all informal physics efforts can be classified as communities of practice. For example, a common type of physics public engagement event is popularized lectures aimed at high school students or adults, such as a "Saturday Physics" series. Often these lectures are given by individual faculty with little extended involvement from university physics students - it is less likely that these efforts could be considered communities of practice.

The programs selected for this study are Science Theatre (based at Michigan State University), PISEC (based at the JILA Physics Frontier Center at the University of Colorado Boulder), and Quavers to Quadratics (a collaboration between University College Dublin, Trinity College Dublin, and the Irish National Concert Hall)[1]. All three programs were initially chosen because they have large numbers of physics student volunteers, have been active for a number of years (Science Theatre since 1991, Quavers to Quadratics since 2014, and PISEC since 2008), and have close links with university physics departments. Furthermore, the design, content, and implementation of the three programs provide a range of formats, from the more common (demonstration show) to the more novel (blending physics and music). They also range in their involvement of physics students, as well as the different types of students, with one group being entirely organized by undergraduate students (Science Theatre), one group being mostly physics graduate students (PISEC), and one group bringing physics students together with students from music (Quavers to Quadratics). Thus these three programs allow us to gain insight into a wide, international variety of informal physics programs that utilize university students as facilitators within the program. Additionally, an important consideration in the selection of programs was that at least one of the authors of this manuscript has been, or is currently, a practitioner associated with each program in some capacity, which provided us with additional insight into the structures and norms of the programs.

[1] You can find more information about Science Theatre at web.pa.msu.edu/sci_theatre, Quavers to Quadratics at www.nch.ie and about PISEC at www.colorado.edu/outreach/pisec.

Below are short descriptions of the format and content of each program in this study. A *domain*, *community,* and the *practices* were identified for each program (more specific details are given in the Appendix). Broadly, the domain of each program was centered around university students achieving a common goal of sharing physics knowledge with youth outside of the field. The specific practices of each program differ considerably since the format of each program is different. However, all three programs involve university physics students taking some form of leading interactive physics activities with youth.

The community of each program consists of all the student volunteers, staff, and faculty who work together to develop and deliver the program content. One significant consideration here is whether to include the youth participants as part of the community as well. In Science Theatre and Quavers to Quadratics, youth typically interact for only a short period of time with the undergraduate facilitators and mostly receive the physics content by listening or touching already developed demonstrations - therefore, we did not include them in the community. However, in PISEC, youth and facilitators meet in small groups weekly over the course of a semester or often an entire year, and they are encouraged to co-create new activities. Due to the extended relationships that are formed in this program, we include youth as part of the PISEC community.

**Quavers to Quadratics** is a program for school children to play with ideas common to physics and music. The children (typically 8 to 12 years of age) are led in their play by undergraduate students from physics, music education, and science education. Quavers to Quadratics is a collaboration between a national cultural institution (the National Concert Hall) and two universities in Ireland (Author Shane Bergin is one of the founders and directors). University students from three departments (physics, science education, and music education) at the corresponding universities participate in the program. One of the aims of Quavers to Quadratics is to have a creative space for undergraduate students to play, work with others from outside their disciplinary niche, and to experience teaching experiences such that they might reflect on their view of physics, music, and education (*domain*). The program runs twice a year with 15-20 undergraduates a semester, who are paid for their effort (*community*). Each three-month cycle requires the undergraduate students to co-plan, co-teach, and co-reflect upon two initial classroom visits at the children's school, a day-long workshop at the National Concert Hall, and a final classroom visit. In these interactions, facilitators encourage children to explore questions like 'what is a wave?' 'how is frequency and amplitude of the wave affects sound', and 'how the shapes and sizes of musical instruments affect sound?' (*practices*).

**Science Theatre** is an undergraduate student group that performs physics and other science demonstrations with a theatrical twist. It is supported by the Department of Physics and Astronomy through monetary contributions and faculty advisement (author Kathleen Hinko has served as a faculty advisor). The majority of students are undergraduates majoring in physics or other STEM fields (*community*). Along with reaching younger audiences (typically 5-17 years old), Science Theatre aims to provide participants with a space to pursue their passions for physics/science and public engagement with like-minded others (*domain)*. Throughout the year, the student organization partners with science events and local schools to perform stage shows and bring hands-on activities for primarily elementary and middle school kids. These demos and activities consist of many physics options (such as a bed-of-nails and a flame tube) as well as topics in chemistry (such as "elephant toothpaste") and various other sciences (*practices*). The group also devotes each spring break to a road trip that aims to visit as many schools as possible in the rural upper peninsula of Michigan. The group has an officer system with twelve elected roles and holds frequent general body meetings.

**PISEC** (Partnerships for Informal Science Education in the Community) is an after-school program, where physics students work with children to explore fun, hands-on physics activities. This program is the main public engagement effort of the JILA Physics Frontier Center for Atomic, Molecular and Optical Physics at the University of Colorado Boulder. (Authors Claudia Fracchiolla and Kathleen Hinko have served as program directors.) The main goal of PISEC is to give children the opportunity to explore pathways to STEM careers (especially physics careers) by engaging them in interactive, inquiry-based physics activities. PISEC also seeks to improve the pedagogical and communication skills of university physics students (*domain*). It runs each university semester with 15-20 mainly physics graduate student volunteers. Student volunteers meet with elementary and middle school children once a week for an hour at the students' school (*community*). During that after-school hour, the children are encouraged to engage in scientific practices through proposing hypotheses, designing their own experiments, and reporting their findings and ideas (*practices*).

By identifying these three essential elements of a community of practice for each informal physics program using a practitioner-researcher approach, we have established that all three cases have the necessary characteristics of a community of practice. This finding is of itself important, as it demonstrates that a diverse range of informal physics programs can serve as an environment for physics students to build identity.

# UNDERSTANDING STUDENT EXPERIENCES IN INFORMAL PHYSICS COMMUNITIES OF PRACTICE

We consider how the domain, community and practices for each group affect student experience, and how in turn, those experiences affect student identity formation. To understand experiences in the informal physics programs, we conducted interviews with students leading these programs (facilitators). Interviews were an appropriate data source because we were looking to explore facilitation experiences through the individual participants' lens alongside their own positionality within their different communities. Interviews were analyzed using the *Community Dimensions* (Table 1) and *Mechanisms of Identity* (Table 2) from the operationalized Community of Practice framework.

## Interviews with university facilitators

A semi-structured interview protocol was designed to ask participants to discuss their experiences in the program and their perceptions toward physics and informal physics programs. Examples of the questions include: Why did you decide to volunteer for the program? Why would you (or would you not) volunteer again? Do you identify as a physicist? How did you end up in physics? The semi-structured interviews allowed us to have the freedom to follow up on questions and be able to capture as much of the narrative as possible. Interviews were conducted by three researchers and lasted about an hour on average.

There were three variations of the protocol, each adapted to the corresponding program. However, all three were centered on questions about the facilitator's motivations for participating in the programs, the expectations and hesitations regarding volunteering, as well as how the students identified themselves regarding their degree. The differences in the protocols consisted mainly of questions pertaining to the particular practices and/or activities of the relevant program. Other differences in the protocols arose from the time when the interview took place. For example, some students were interviewed once - after they had participated in the program for at least one semester, while other students were interviewed twice during their participation in a program. This difference in frequency and timing of the interviews was mainly due to the nature of the programs. For example, two of the programs require a semester long commitment (PISEC and Quavers to Quadratics) whereas the third program (Science Theatre) allows for students to participate when time allows. Also, two of the programs, Quavers to Quadratics and Science Theatre, offer a week-long trip during spring break or summer break. In the event that one student had completed two or more interviews, we viewed the collection of interviews as a single snapshot of the student's relationship with the program.

We collected interviews from 58 participants across the three informal physics programs. These participants were from the pool of those facilitating the activities in the programs and that were willing to participate in the research study. For this paper we focused on a subset of 18 participants, which correspond to 29 total interviews. Seven participants were from PISEC (7 interviews), five participants were from Science Theatre (10 interviews), and six participants were from Quavers to Quadratics (12 interviews). There are several justifications for using a subset of the full data set for analysis: 1) Consistency of interview protocols - earlier versions of the PISEC protocol differed from the protocols used for the other two programs, because we used PISEC participants to test the different protocols. We conducted three different test rounds of the protocol before arriving at the final version, which is the one used in this study and 2) interviews used in earlier stages to operationalize and validate the framework [12] were excluded from the current study. Each interview within our entire data set was read, or listened to, by members of the research team and the final interviews selected for this paper were intentionally chosen to represent the diversity of the facilitator population within the programs.

The interviews used for this paper were conducted by three female researchers: one white woman and two women of color. Interviews were conducted locally, that is, the researcher conducting the interview was, at the time of the interview, based in the institution where the program is hosted. The large majority of the interviews were conducted in person using audio recordings. The length of the interviews varied between 30 minutes to an hour, depending on how much detail the interviewee gave in the questions and some of the variations in the protocol.

## Coding the interviews

In order to understand what structural elements of each informal physics program contributed to fostering a physics identity among the university student facilitators, we used our operationalized Communities of Practice framework to analyze the interviews [12]. During the coding process, we used the *Community Dimension* (first layer of coding) codes to establish membership and the *Mechanisms of Identity* (second layer of coding) to determine the agents of change for the membership levels. To understand levels of membership within the community, we used our first layer of coding and assigned one of the five subcodes (*insider*, *inbound*, *neutral*, *outbound*, *peripheral*) to any portion of the interview coded with a *Community Dimension* code. The subcodes denote position within membership levels. This position can be static, in the cases of *insider* or *peripheral*, or dynamic in the cases of *inbound* or *outbound*. A *neutral* code indicated that from the coded section we could not determine present membership levels. For example, if a portion of the interview was coded as *Accountability to the Enterprise* and the interviewee was describing how their understanding of the community's mission was evolving, then that same section of the interview would also be coded with the *inbound* subcode.

Additionally, the second layer of coding was used to identify the *Mechanisms of Identity*. The subcodes for the *Mechanisms of Identity* indicate whether the particular mechanism being coded prompted a movement inward or outward in the levels of membership. In some instances, the same interview segment could be coded with *Community Dimension* and *Mechanism of Identity,* however a coded segment cannot be assigned two or more codes from the same layer (i.e. the same portion of the interview could not be assigned *Accountability to the Enterprise* and *Mutuality of Engagement).* In later sections we denote which code was assigned to each quote that is shared in this paper.

The third and final layer of coding included assigning the community codes. These codes showed which community or communities of practice were being discussed by the interviewee at that instance of the interview. The list of communities discussed throughout this data set included the three informal programs, the student's discipline (usually the physics community), and the graduate school community. Any quotes shared in this paper were coded with one of the program communities and often other communities as the facilitators' experiences in one community could impact their participation in another community.

## Coding validation process

The interview coding process required many members of the research team and was carried out in two main phases: a training and validation phase (where only a few interviews were coded) and an independent coding phase (where the remaining interviews were coded). This two phase approach allowed us to ensure that each member of the research team understood the framework and felt comfortable coding the interviews. In addition, this phased approach allowed us to utilize inter-rater reliability to ensure that the coding was consistent among all of the interviews and that the codebook could be used by other researchers.

The first phase of the coding process was focused on validation of the codebook and training each member of the research team. First, we had two members of the research team independently code an interview for one program with a set of codes (*Community Dimensions* or *Mechanisms of Identity*) without the subcodes or communities. Then, the researchers compared their coding in order to discuss and reconcile any differences. This part of the process was meant to check how every part of the interview was coded, not only which codes were used but also the length of the coded segments. If there was a discrepancy about the code used, then there would be a discussion in which each researcher would explain why the corresponding segment was assigned the particular code. Throughout this process, we updated and created a robust codebook with our operationalized framework. Once each of the researchers had a good understanding of the coding framework, a second interview was coded independently. The researchers then compared the codes and estimated kappa value for the interrater agreement value. Once agreement of over 80% had been reached, then the researchers independently coded the interviews with the corresponding subcodes and communities. The same process for coder agreement was conducted for each program and for each set of researchers.

The second phase of coding was planned to ensure that all interviews were coded and that there was discussion between the researchers during this process. First, we split into teams so that all six of the researchers coding would only focus on one set of constructs (i.e. each researcher only coded for the *Community Dimensions* or the *Mechanisms of Identity* during this phase). Each researcher was in charge of a set of interviews for coding with the corresponding set of codes and independently coded those interviews. Their partner would then independently code one of the interviews in the set with the corresponding codes to check for interrater reliability. If differences arose, those were discussed with an external researcher member overviewing the process, and the differences were reconciled. Once agreement was more than 80% then the rest of the coded interviews are reviewed by the partner without having to code independently the interview. This team approach allowed us to ensure that all of the interviews were coded in a timely manner and that interrater reliability was checked for each researcher.

### Statistical analysis methods

To analyze the data to compare the overall programs based upon individual viewpoints, the counts (frequency) of all of the codes from each interview were normalized, i.e. we used percentages to report findings. For each interview, the percentage of each of the individual code is given with respect to the total number of codes in the interview. For example, in order to calculate the percentage of *Accountability to the Enterprise* for an interviewee from Quavers to Quadratics, we divided the frequency of *Accountability to the Enterprise* codes for the total frequency of *Community Dimensions* codes within that interview. It is important to note here that for the purpose of this study we only looked at the codes that were associated with the program's community of practice, i.e. for interviewees from Quavers to Quadratic we only looked at the *Community Dimension* and *Mechanisms of Identity* codes that were coded under the Quavers to Quadratics community. More details about the statistics reported, including what statistical tests were conducted can be found in the appendix.

## INTERVIEW ANALYSIS

We looked at the coded interviews from student facilitators within the three programs, coupled with our existing practitioner-based knowledge, to understand how structural similarities and differences between the programs impacted these student facilitators. In order to do this, we look at the *Community Dimensions* and *Mechanisms of Identity* codes for each program. We also look at the overlap between these layers of codes; this intersection provides information on whether facilitators' membership within their respective community of practice changed throughout their participation in the community and what aspects of the corresponding program's design may have prompted such changes.

## *Community Dimensions*

From the overall *Community Dimension* codes, we can understand how students perceived their positionality within the programs as well as which dimensions impact their positionality. Figure 2 shows the *Community Dimensions* aggregated for each of the three informal programs. One finding to note, is there are no statistically significant differences between any of the dimensions for Quavers to Quadratics and Science Theatre. In the case of these two programs, *Negotiability of the Repertoire* - which is related to knowledge of and competence in the practices of the community - represents more than 40% of the *Community Dimension* codes for both programs, making it the most important factor to establish membership, closely followed by *Accountability to the Enterprise*, which represents more than 35% of the codes. We interpret this to mean that facilitators in Quavers to Quadratics and Science Theatre understand that to become a central member of the community, they need to actively participate in the practices of the community (*Negotiability of the Repertoire*) and have a clear understanding of the mission of the community (*Accountability to the Enterprise*). In contrast, *Mutuality of Engagement* represented less than 20% of the codes for these programs, indicating that connecting to other program facilitators is less important for students in these groups. This interpretation of student experiences is consistent with other theories in which competence, performance, and connection to personal interests are among the main constructs for identity development [14,18]. The main difference in the Community Dimension codes between these two programs is seen in the subcode percent (see table in Figure 2). Science Theatre facilitators had higher percentages of neutral codes and Quavers to Quadratics had more *insider* and *inbound* codes across all categories. More *insider* and *inbound* codes for Quavers to Quadratics indicate that the facilitators perceive more strongly that they are central members of that community, or are moving to be more central.

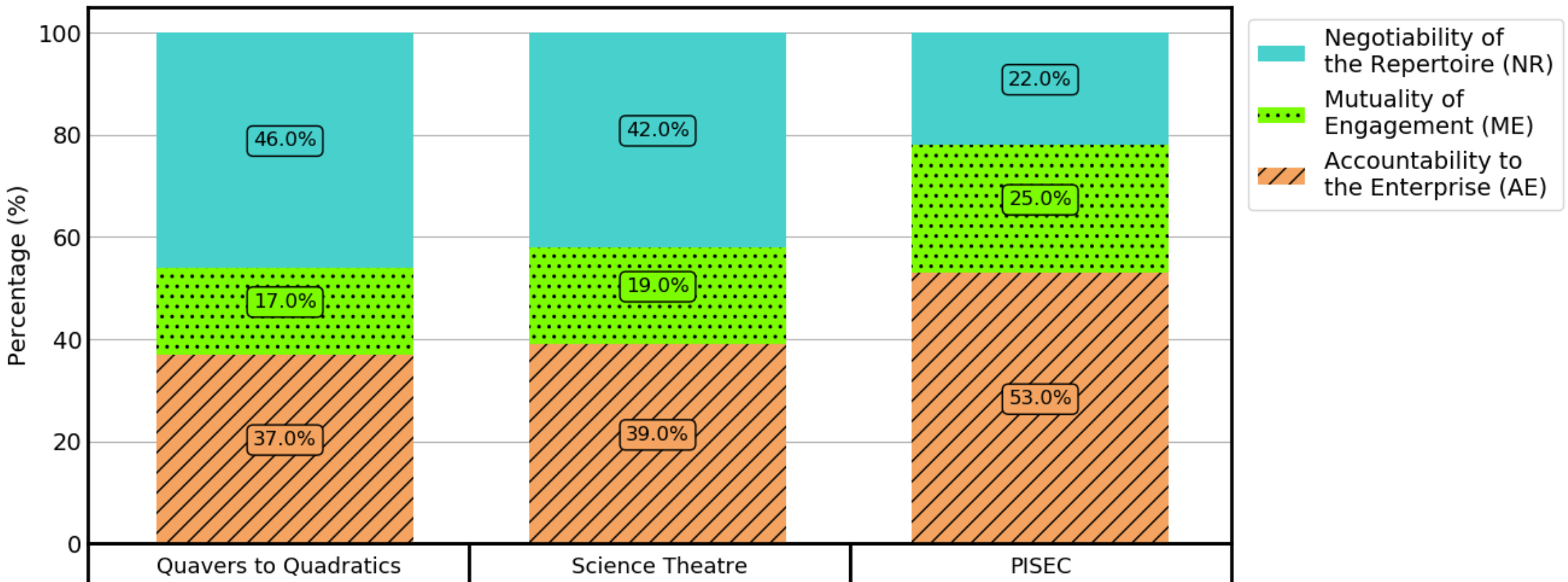

| | Quavers to Quadratics | | | Science Theatre | | | PISEC | | |
|---|---|---|---|---|---|---|---|---|---|
| | AE | ME | NR | AE | ME | NR | AE | ME | NR |
| Insider | 55.5 | 20.9 | 38.8 | 14.0 | 1.4 | 13.0 | 13.1 | 5.0 | 0.0 |
| Inbound | 28.2 | 72.1 | 16.5 | 53.0 | 54.8 | 26.0 | 57.1 | 60.0 | 51.4 |
| Neutral | 9.1 | 7.0 | 27.3 | 27.0 | 39.7 | 49.0 | 7.1 | 20.0 | 22.9 |
| Peripheral | 6.4 | 0.0 | 5.8 | 2.0 | 1.4 | 0.6 | 17.9 | 0.0 | 8.6 |
| Outbound | 0.9 | 0.0 | 11.5 | 4.0 | 2.7 | 11.0 | 4.8 | 15.0 | 17.1 |

**Figure 2:** The above graph shows the overall *Community Dimension* codes for each of the three informal programs (left to right: Quavers to Quadratics, Science Theatre, PISEC). The percentages shown in each bar indicate how many

of the codes were assigned to each *Community Dimension* out of the total *Community Dimensions* codes per interview, and thus all the bars add up to 100%. The table shows the breakdown of each *Community Dimension* into the subcodes with the percentage indicating how often the subcode was assigned to a given *Community Dimension.*

In contrast to the other two informal programs, for PISEC, *Accountability to the Enterprise* makes up 53% overall of the *Community Dimension* codes. Thus, PISEC facilitators perceived the commitment to the goals (*domain*) of the community as the leading factor for building membership. The majority of the facilitators had a clear alignment with the values of the program and were demonstrating a strong commitment to the program, shown by a majority of *inbound* codes within this dimension. *Mutuality of Engagement* makes up 25% of the codes for PISEC, indicating that connecting to other program facilitators is a more important factor for students in PISEC than in the other two programs. *Negotiability of the Repertoire* is the least frequent code (22%), which shows that PISEC facilitators view understanding the practices as the least important aspect of being a member of the group. Reasons for these differences are due to the different structures of the PISEC program and who constitutes the community. These will be expanded on in the next section.

### *Mechanisms of Identity*

While the *Community Dimensions* indicate facilitators' perceived level of membership within the community of practice, the *Mechanisms of Identity* point out how the particular structures and practices of the program might foster movement in perceived membership levels. Figure 3 shows the *Mechanisms of Identity* code distribution for each informal program. Overall, we observe that participation in informal physics programs has a positive effect on facilitators' identities, as shown by the high percentages of *inbound* codes across all three programs (see first row of table in Figure 3 for details). Furthermore, across all three programs *Negotiated Experiences* and *Community Membership* are the most salient mechanisms, suggesting interactions and connections made through participation while engaging in the practices of the community support membership and identity development. This finding is consistent with other research in which recognition (both internally and through external validation) as a member of a community of practice can be a determinative factor for identity formation [7, 11, 12]. Differences arise among the programs when we look at the *Mechanisms of Identity* (as compared to the *Community Dimensions*). While all three of the programs have similar missions (*domains*), they each have a unique way of carrying out this mission. This variety of program structures is reflected among the *Mechanisms of Identity* distributions, which will be discussed in detail in the following sections.

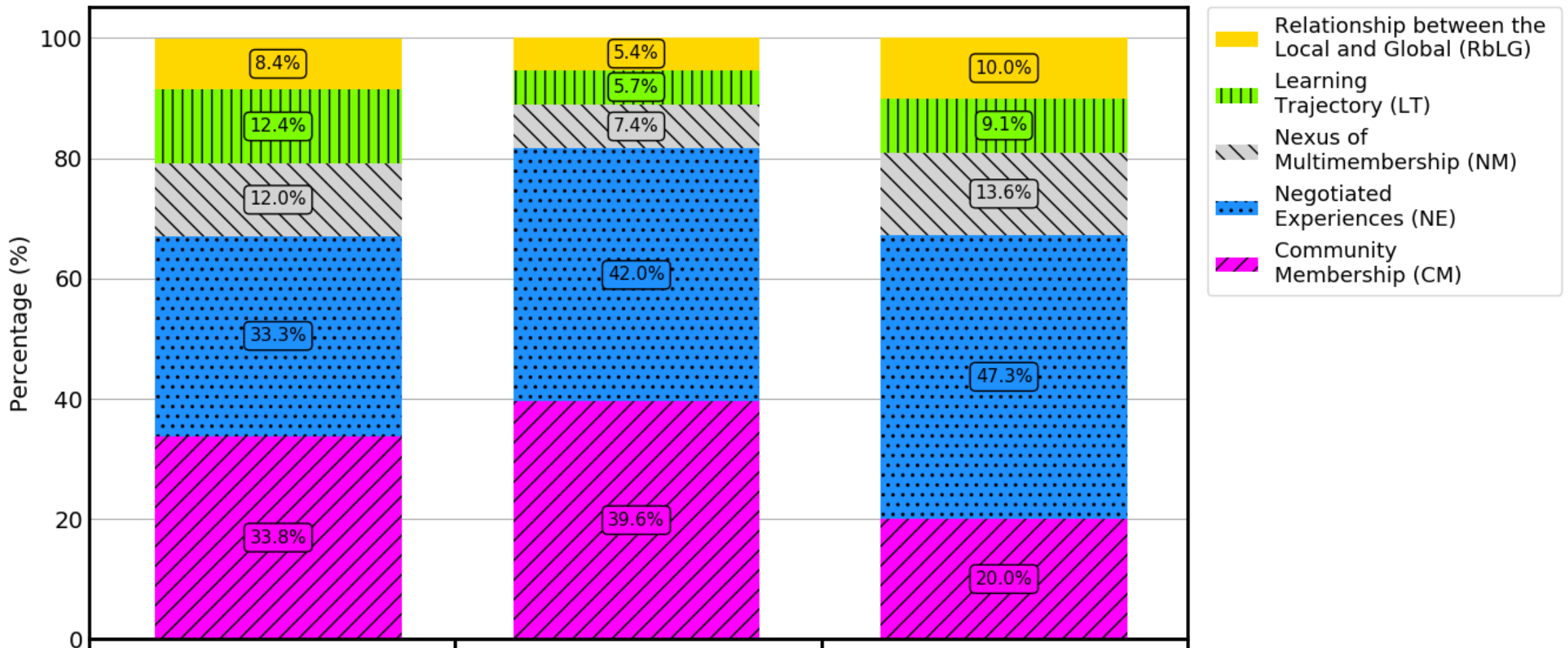

| | Quavers to Quadratics | | | | | Science Theatre | | | | | PISEC | | | | |
|---|---|---|---|---|---|---|---|---|---|---|---|---|---|---|---|
| | CM | NE | NM | LT | RbLG | CM | NE | NM | LT | RbLG | CM | NE | NM | LT | RbLG |
| Inbound | 60.0 | 80.0 | 81.0 | 57.1 | 94.7 | 50.4 | 72.3 | 52.0 | 84.2 | 88.9 | 68.2 | 64.7 | 33.3 | 70.0 | 63.6 |
| Neutral | 32.0 | 19.0 | 15.0 | 42.9 | 5.3 | 40.6 | 19.1 | 36.0 | 15.8 | 5.6 | 22.7 | 23.4 | 40.0 | 30.0 | 36.4 |
| Outbound | 8.0 | 1.0 | 4.0 | 0.0 | 0.0 | 9.0 | 8.5 | 12.0 | 0.0 | 5.6 | 9.1 | 11.8 | 26.7 | 0.0 | 0.0 |

**Figure 3:** The above graph shows the overall *Mechanisms of Identity* codes for each of the three informal programs (left to right: Quavers to Quadratics, Science Theatre, PISEC). The percentages shown in each bar indicate how many of the codes were assigned to each *Mechanism of Identity* out of all *Mechanisms of Identity*, and thus all the bars add up to 100%. The table shows the breakdown of each *Mechanism of Identity* into the subcodes with the percentage indicating how often the subcode was assigned to a given *Mechanism of Identity*.

# INTERSECTIONS BETWEEN COMMUNITY DIMENSIONS AND MECHANISMS OF IDENTITY

In order to gain a more thorough understanding of facilitator experiences within each program, we now look at the interactions of the *Mechanisms of Identity* with the *Community Dimension* codes. These intersections give insight as to which structural elements within the programs impact facilitator perception of their membership and their identity.

To map the intersections, we looked at the overlap in counts between the Mechanisms of Identity and each Community Dimension. We then graphed these values on a pseudo-scale that divides the interaction between *Mechanisms of Identity* subcodes and one of the *Community Dimensions* into four quadrants (see Figures 4-6). The X-axis has subcodes for the *Community Dimensions* going from left to right, ranging from *outsider* to *insider*. The Y-axis scale was determined by the *Mechanisms of Identity*, ranging from *outbound* to *inbound,* going from bottom to top. (See the Appendix for more details). Different shaped markers were used for each of the mechanisms, with the size of the markers representing the normalized frequency (percentage) of the interaction. Plotting the data in this way allows us to visually see which mechanisms are most prevalent with respect to inbound/insider community of practice experiences (and likewise outbound/outsider experiences). For example, the markers in the upper right quadrant (region shaded green on the table) indicate *Mechanisms of Identity-inbound*/*Community Dimensions-insider* interactions and markers in the lower left quadrant (region shaded red on the table) show *Mechanisms of Identity-outbound*/*Community Dimension-peripheral* interactions. Below, we discuss the key interactions (seen by the larger markers in each graph) for each of the three *Community Dimensions* through quotes from the facilitators (using pseudonyms) to narrate which aspects of program design impacted the facilitator's identity. These observations will also be supported by the author's practitioner knowledge of the programs in order to provide a complete picture.

## Intersection with *Accountability to the Enterprise*

*Accountability to the Enterprise* is related to commitment and understanding of the mission of the program and how facilitators align that mission with their personal values. *Accountability to the Enterprise* represents more than 35% of the *Community Dimension* codes across the three programs (see Figure 2 for details). This suggests that those who chose to facilitate informal physics programs have a level of understanding and are in alignment with the values of public engagement and communicating physics and science more generally. They see the value in the work that is done through these programs and believe it is important. We saw this in an interview with Levi, a physics grad student volunteering with PISEC, when he says:

> I guess, like, as an adult being in PISEC and other outreach programs it helps me, like, see directly the importance of outreach and these types of, like, connections in the community for scientists. I think I always, like, understood why these programs were important, but maybe not, not as fully as I do now. Being a physicist and a physics education researcher and also a participant in these programs, I feel like I have a better grasp on, like, why this is so important to do [sic].

We consistently saw that the connection with the *domain* (mission) of the community was a significant aspect of a facilitator's choice to participate in an informal physics education program.

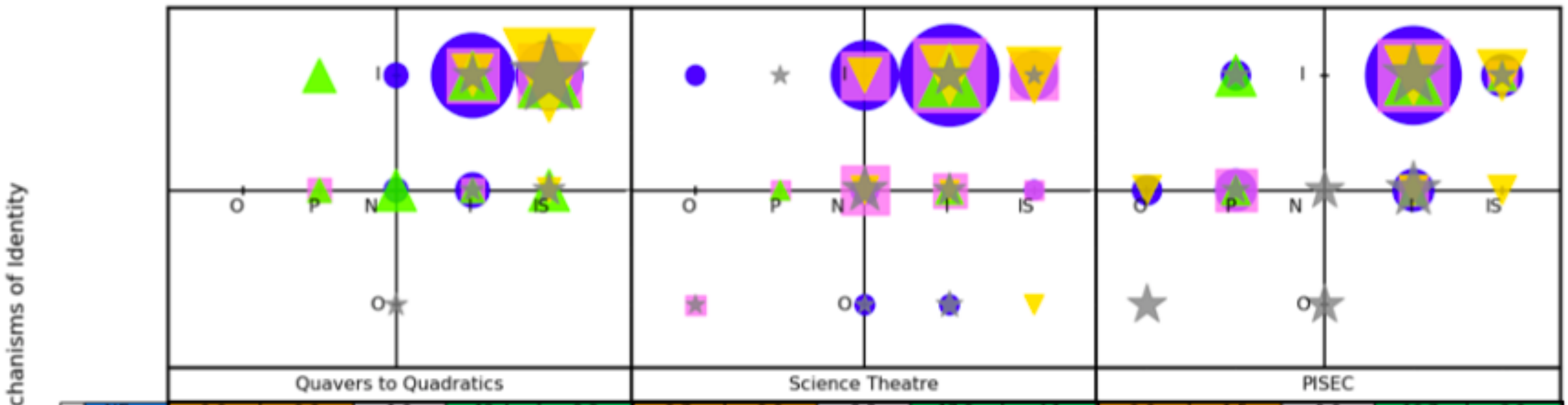

| | | Quavers to Quadratics | | | | | Science Theatre | | | | | PISEC | | | | |
|---|---|---|---|---|---|---|---|---|---|---|---|---|---|---|---|---|
| | | Outbound | Peripheral | Neutral | Inbound | Insider | Outbound | Peripheral | Neutral | Inbound | Insider | Outbound | Peripheral | Neutral | Inbound | Insider |
| Inbound | NE o | 0.0 | 0.0 | 1.0 | 12.4 | 8.3 | 0.7 | 0.0 | 8.3 | 17.9 | 4.1 | 0.0 | 1.5 | 0.0 | 16.7 | 3.0 |
| | CM □ | 0.0 | 0.0 | 0.0 | 5.2 | 7.2 | 0.0 | 0.0 | 4.1 | 9.7 | 4.1 | 0.0 | 0.0 | 0.0 | 9.1 | 1.5 |
| | LT △ | 0.0 | 2.1 | 0.0 | 4.1 | 7.2 | 0.0 | 0.0 | 0.0 | 6.9 | 0.0 | 0.0 | 3.0 | 0.0 | 6.1 | 1.5 |
| | RbLG ▽ | 0.0 | 0.0 | 0.0 | 3.1 | 15.5 | 0.0 | 0.0 | 2.1 | 6.2 | 5.5 | 0.0 | 0.0 | 0.0 | 6.1 | 4.6 |
| | NM ☆ | 0.0 | 0.0 | 0.0 | 3.1 | 12.4 | 0.0 | 0.7 | 0.0 | 3.5 | 0.7 | 0.0 | 1.5 | 0.0 | 7.6 | 1.5 |
| Neutral | NE o | 0.0 | 0.0 | 1.0 | 2.1 | 0.0 | 0.0 | 0.0 | 1.4 | 0.0 | 0.7 | 1.5 | 3.0 | 0.0 | 3.0 | 0.0 |
| | CM □ | 0.0 | 1.0 | 0.0 | 1.0 | 0.0 | 0.0 | 0.7 | 4.1 | 2.1 | 0.7 | 0.0 | 3.0 | 0.0 | 1.5 | 0.0 |
| | LT △ | 0.0 | 1.0 | 3.1 | 1.0 | 3.1 | 0.0 | 0.7 | 0.0 | 1.4 | 0.0 | 0.0 | 1.5 | 0.0 | 1.5 | 0.0 |
| | RbLG ▽ | 0.0 | 0.0 | 0.0 | 0.0 | 1.0 | 0.0 | 0.0 | 1.4 | 0.7 | 0.0 | 1.5 | 0.0 | 0.0 | 1.5 | 1.5 |
| | NM ☆ | 0.0 | 0.0 | 0.0 | 1.0 | 2.1 | 0.0 | 0.0 | 4.1 | 2.1 | 0.0 | 0.0 | 1.5 | 3.0 | 6.1 | 0.0 |
| Outbound | NE o | 0.0 | 0.0 | 0.0 | 0.0 | 0.0 | 0.0 | 0.0 | 0.7 | 0.7 | 0.0 | 0.0 | 0.0 | 0.0 | 0.0 | 0.0 |
| | CM □ | 0.0 | 0.0 | 0.0 | 0.0 | 0.0 | 0.7 | 0.0 | 0.0 | 0.0 | 0.0 | 0.0 | 0.0 | 0.0 | 0.0 | 0.0 |
| | LT △ | 0.0 | 0.0 | 0.0 | 0.0 | 0.0 | 0.0 | 0.0 | 0.0 | 0.0 | 0.0 | 0.0 | 0.0 | 0.0 | 0.0 | 0.0 |
| | RbLG ▽ | 0.0 | 0.0 | 0.0 | 0.0 | 0.0 | 0.0 | 0.0 | 0.0 | 0.0 | 0.7 | 0.0 | 0.0 | 0.0 | 0.0 | 0.0 |
| | NM ☆ | 0.0 | 0.0 | 1.0 | 0.0 | 0.0 | 0.7 | 0.0 | 0.7 | 1.4 | 0.0 | 3.0 | 0.0 | 3.0 | 0.0 | 0.0 |

Accountability to the Enterprise

**Figure 4**: The above figure shows the intersection of the *Accountability to the Enterprise Community Dimension* subcodes (x axis) with all five of the *Mechanisms of Identity* and their subcodes (y axis). Each mechanism is represented by a different color and shape as shown in the second column of the table. The frequencies of the overlaps are shown by the size of the shape and the exact percentage in the table below which follows the same layout as the graph with the four quadrants colored (green: *inbound/insider* with *inbound*, orange *inbound/insider* with *outbound* or *outbound/peripheral* with *inbound*, red: *outbound/peripheral* with *outbound*).

Figure 4 shows how *Accountability to the Enterprise* intersects with the all five of the *Mechanisms of Identity* constructs in each of the three programs. Having all five *Mechanisms of Identity* interact with the *Accountability to the Enterprise* demonstrates that facilitators see an alignment of their values and commitment to the mission of the program and how that impacts all areas of their membership. The high level of commitment from the facilitators is seen through the high percentage of the *inbound* and *insider* subcodes within the *Accountability to the Enterprise* code (more than 67% for all programs) shown in the upper-right quadrant of Figure 4. However, we also see that *Community Membership*, *Learning Trajectory*, and *Relationship between Local and Global* are the mechanisms that interact with this community dimension most frequently. We discuss these interactions in the context of the programs below (with more information about the other mechanisms in the appendix).

***Negotiated Experiences*** (shown by the blue circle in Figure 4) is seen frequently interacting with *Accountability to the Enterprise* and accounts for the majority of the inward movement in membership among facilitators.

In PISEC, building long-term relationships with the children and seeing the effect of outreach on their lives made the facilitators "hopeful" and really enjoy the experience. For the PISEC facilitators, the connection with children seems to be more impactful than those with other peers, community partners, or other members of the community. For example, Lily discusses how interacting with the students increases her excitement about physics:

> I mean I love physics and I love science, but it sort of makes me even more excited about it, being there, like working on it with the kids and kind of like talking about physics concepts in really simple levels kind of makes it more exciting to me, like 'wow this stuff is awesome.' So in that sense I would say they had an impact on me [sic].

This was coded as *inbound Accountability to the Enterprise* and *inbound Negotiated Experiences* because by engaging children in physics and demonstrating to them that physics is fun (part of PISEC domain), Lily is at the same time reiterating her passion for physics.

Similarly, for Science Theatre the common themes from the interactions of *Accountability to the Enterprise* and *Negotiated Experiences* are related to how the facilitators enjoyed the children being really engaged, asking questions, getting involved, and expressing that they think science is cool. Tom shared about his experience with students reacting to the demos:

> And then, being able to share the science to little kids is really awesome, 'cause I love science, so [laughter] having the idea that I might inspire a little, or a kid, to even just think something that I did was cool, that's science related [sic].

This was coded as *inbound Accountability to the Enterprise* and *inbound Negotiated Experiences* because, like Lily from PISEC, Tom enjoys the children's positive reaction to the shows (*Negotiated Experiences - inbound*) and the idea that he might be inspiring children to continue participating in physics or science (Science Theatre *domain*). This makes participation in the Science Theatre community valuable to him and therefore supports inward movement to the community. Additionally, it reaffirms his passion for science - supporting his science identity.

The engagement and excitement children show while engaging in the activities is a source of motivation and inspiration for many of the Quavers to Quadratics facilitators. Emer, a science education major, discussed her thoughts about work with the children through the program:

> I definitely feel like I'm making some sort of meaningful connections which, like, is such a great thing to see because that's obviously the aim of it... And when I see the kids thinking: 'Oh my God this is amazing! I didn't know or think of it this way!' or you know like seeing them, they connect vibrations and amplitude

> together, and those two things [are] both scientific… You can see that they are more and more convinced each time [about the connections between physics and music] [sic].

This reflection, coded as *insider Accountability to the Enterprise* and *inbound Negotiated Experiences,* on how enjoyable it is to see children at the end of the workshop connecting the concepts of music and physics (Quavers to Quadratics *domain*) translates into meaningful experiences for the facilitator, therefore fostering inward movement in their membership.

***Community Membership*** is the second most coded *Mechanism of Identity* (shown by the pink square in Figure 4) and tracks how facilitators' confidence in the practices of the community impact their membership. This identity mechanism is related to elements of the structural design of the program; in the case of Science Theatre, this would be related to the community built collection of demonstrations and show scripts along with the physical spaces that the group has to hold meetings. Science Theatre facilitators are undergraduate students from various different backgrounds and areas of study including both science and non-science majors. Therefore, they have a considerable amount of content to master, as well as needing to learn how the demonstrations work and the phenomena behind them. For example, Liam, whose background is not in physics, stated that he feels like a science person because he asks a lot of questions:

> But I was asking questions like, this dixie cup demo worked really really well. Does it have to be paired with the bed of nails? Could you do bed of nails with dixie cups and then have another part, just have the dixie cups demo. Because you could do that. That demo in and of itself is pretty great… I was asking a lot of practical questions where I'm like, in practicality, how can we make this work better? How can we explain this to where it makes more sense [sic]?

In this quote, Liam is explaining that he is insistent in learning the content and practices of the program by asking his peers (*Community Membership*). Liam's questions show his growing knowledge of and confidence with the Science Theatre practices so this segment was coded as *inbound Accountability to the Enterprise* and *inbound Community Membership*. He goes on to explain that he cares more about making the experience better for the children and assuring that they do not leave with misconceptions even if they do not remember the concepts. He does this because he cares about the experiences the children have in the program and making science fun and understandable (part of the domain of Science Theatre).

***Relationship between Local and Global*** (shown by the yellow triangle) and the *insider* subcode for *Accountability to the Enterprise* also frequently, as seen in Figure 4. This mechanism relates to the connection between participation in the local community and participation in a broader (global) community of practice. Developing awareness of those connections requires a deep understanding of the domain of the local community and how that plays a role into a more global perspective.

Across the programs the main theme identified in segments coded as *Accountability to the Enterprise* relates to the facilitators' wish to be part of a community focused on encouraging children to enter STEM fields. For many interviewees, being involved with Science Theatre provided a bit more than just a way to satisfy their physics curiosity, it provided a chance to share physics with others. One interviewee, Grace, connects her interest and involvement in Science Theatre to her other identities and shares how this intersection is important to her:

> I really like [Science Theatre] because it's like a chance to reach out and spread what you're passionate about, and I think that, especially as a woman of color, I want to see more people in science in general. Especially reaching out to young girls, young girls of color. Because, you know, science can be very monolithic and when you do outreach you get more types of people [sic].

In this case, Grace's statement was coded as *insider Accountability to the Enterprise* because she is connecting her value of wanting to be a role model for women of color in physics within Science Theatre's mission to provide fun and engaging physics experiences for everyone. This portion of Grace's interview was also coded as *inbound Relationship between Local and Global* because she is describing how her participation in Science Theatre can have a broader impact on the physics/science community at large, regarding issues of underrepresentation.

In the case of PISEC, the facilitators are already committed to the physics community and its value, given that they are mostly PhD students, so what PISEC allows them to do is communicate their passion for physics to children. As Evan explains:

> I've kind of, like, really grown to like I would do PISEC now even if my boss told me I didn't have to, just because I feel like it's a great idea to give back to the community and to kind of just be there as someone who can say 'hey, science is cool.' Because I didn't have anybody when I was that age telling me science was cool, so I feel like taking just an hour and a half out of your time, out of your day, and kind of just introducing the world of science to some kids is a good thing, is a fun thing to do [sic].

This portion of the interview was coded as *insider Accountability to the Enterprise* and *inbound Relationship between Local and Global* because there is an indication of a commitment to participate in the community, regardless of the implications and therefore perceiving the value of doing outreach. Evan is reflecting on how he would have liked the opportunity to participate in a program like PISEC as a child, and that is something that he can contribute now.

In Quavers to Quadratics, student interviews show that the facilitators hold similar views to their PISEC and Science Theatre counterparts around the importance of the program's values, as said by Seán:

> [...] I mean I have always been a big believer in outreach and I think it's very important especially for scientists to do. Maybe I am bias because I am involved in science. But it's affected how I view it and I think I can put more emotion to it now because I had taken part in it... But I have always been a big believer in outreach so it didn't change that in any way shape or form but yeah I understand now there is more emotion behind that and how gratifying it can be [sic].

This quote from Seán was coded as *insider Accountability to the Enterprise* and *inbound Relationship between Local and Global* because he is explaining how he sees the value of the outreach community and how that intersects with the domain of the physics/science community. Sean is also narrating how participation in Quavers to Quadratics has reaffirmed the positive feelings related to being a member of the science outreach community.

## Intersection with *Mutuality of Engagement*

Across all three programs, as seen in Figure 2, *Mutually of Engagement* represents no more than 25% of the *Community Dimension* codes. However, at the subcodes level (see table in Figure 2) we see that relationships with other members fosters and supports membership, as demonstrated by the majority of the codes (>50% for all programs) falling into the *insider/inbound* levels.

*Mutuality of Engagement* is related to interactions between community members, the norms of those interactions, and how we recognize and are recognized as members of the community of practice. Given the definition of *Mutuality of Engagement,* it is expected that this *Community Dimension* would have more interaction with the *Mechanisms of Identity* related to interactions with other members of the community such as *Negotiated Experiences* and *Community Membership*. This is confirmed in Figure 5, where the majority of the interactions for all three programs are associated with these *Mechanisms of Identity*.

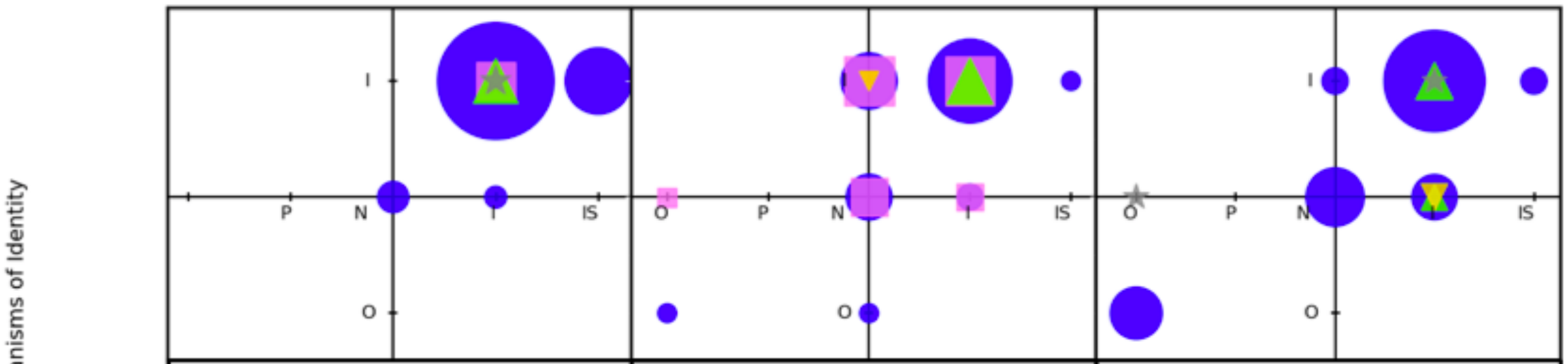

| | | Quavers to Quadratics | | | | | Science Theatre | | | | | PISEC | | | | |
|---|---|---|---|---|---|---|---|---|---|---|---|---|---|---|---|---|
| | | Outbound | Peripheral | Neutral | Inbound | Insider | Outbound | Peripheral | Neutral | Inbound | Insider | Outbound | Peripheral | Neutral | Inbound | Insider |
| Inbound | NE o | 0.0 | 0.0 | 0.0 | 57.1 | 18.4 | 0.0 | 0.0 | 13.2 | 29.4 | 1.5 | 0.0 | 0.0 | 2.9 | 42.9 | 2.9 |
| | CM □ | 0.0 | 0.0 | 0.0 | 6.1 | 0.0 | 0.0 | 0.0 | 10.3 | 10.3 | 0.0 | 0.0 | 0.0 | 0.0 | 0.0 | 0.0 |
| | LT △ | 0.0 | 0.0 | 0.0 | 8.2 | 0.0 | 0.0 | 0.0 | 0.0 | 8.8 | 0.0 | 0.0 | 0.0 | 0.0 | 5.7 | 0.0 |
| | RbLG ▽ | 0.0 | 0.0 | 0.0 | 0.0 | 0.0 | 0.0 | 0.0 | 1.5 | 0.0 | 0.0 | 0.0 | 0.0 | 0.0 | 0.0 | 0.0 |
| | NM ☆ | 0.0 | 0.0 | 0.0 | 4.1 | 0.0 | 0.0 | 0.0 | 0.0 | 0.0 | 0.0 | 0.0 | 0.0 | 0.0 | 2.9 | 0.0 |
| Neutral | NE o | 0.0 | 0.0 | 4.1 | 2.0 | 0.0 | 0.0 | 0.0 | 8.8 | 2.9 | 0.0 | 0.0 | 0.0 | 14.3 | 8.6 | 0.0 |
| | CM □ | 0.0 | 0.0 | 0.0 | 0.0 | 0.0 | 1.5 | 0.0 | 5.9 | 2.9 | 0.0 | 0.0 | 0.0 | 0.0 | 0.0 | 0.0 |
| | LT △ | 0.0 | 0.0 | 0.0 | 0.0 | 0.0 | 0.0 | 0.0 | 0.0 | 0.0 | 0.0 | 0.0 | 0.0 | 0.0 | 2.9 | 0.0 |
| | RbLG ▽ | 0.0 | 0.0 | 0.0 | 0.0 | 0.0 | 0.0 | 0.0 | 0.0 | 0.0 | 0.0 | 0.0 | 0.0 | 0.0 | 2.9 | 0.0 |
| | NM ☆ | 0.0 | 0.0 | 0.0 | 0.0 | 0.0 | 0.0 | 0.0 | 0.0 | 0.0 | 0.0 | 2.9 | 0.0 | 0.0 | 0.0 | 0.0 |
| Outbound | NE o | 0.0 | 0.0 | 0.0 | 0.0 | 0.0 | 1.5 | 0.0 | 1.5 | 0.0 | 0.0 | 11.4 | 0.0 | 0.0 | 0.0 | 0.0 |
| | CM □ | 0.0 | 0.0 | 0.0 | 0.0 | 0.0 | 0.0 | 0.0 | 0.0 | 0.0 | 0.0 | 0.0 | 0.0 | 0.0 | 0.0 | 0.0 |
| | LT △ | 0.0 | 0.0 | 0.0 | 0.0 | 0.0 | 0.0 | 0.0 | 0.0 | 0.0 | 0.0 | 0.0 | 0.0 | 0.0 | 0.0 | 0.0 |
| | RbLG ▽ | 0.0 | 0.0 | 0.0 | 0.0 | 0.0 | 0.0 | 0.0 | 0.0 | 0.0 | 0.0 | 0.0 | 0.0 | 0.0 | 0.0 | 0.0 |
| | NM ☆ | 0.0 | 0.0 | 0.0 | 0.0 | 0.0 | 0.0 | 0.0 | 0.0 | 0.0 | 0.0 | 0.0 | 0.0 | 0.0 | 0.0 | 0.0 |

Mutuality of Engagement

**Figure 6:** The above figure shows how the *Mechanisms of Identity* overlap with the *Mutuality of Engagement*. The frequency of the overlapping is represented by the size of the markers. The table mirrors the plot and shows the frequencies in each quadrant marked with different colors

***Negotiated Experiences*** (shown by the blue circle in Figure 6) represents one of the most important mechanisms to build identity within the communities of practices observed here. The relationships with other members of the community, whether the children, partners, or peers, fosters and supports facilitators' membership. Feeling supported and recognized by their peers provided a steppingstone for newer members to venture into becoming more central members. The excitement and engagement children showed towards physics was a contagious feeling that reminded facilitators of their passion for physics and encouraged them to continue participating and become more central members of the community. Additionally, *recognition* - being recognized as a member of the community has long been proved to be one of the main aspects of building identity [14]. This is true for the programs' community of practice as well as the discipline's community of practice. Through participation in the informal physics communities, children and peers recognize the facilitators as members of the programs' community of practice and as experts of the discipline community of practice.

Moreover, the majority of the *Mutuality of Engagement-Negotiated Experiences* interactions are on the upper-right quadrant, meaning that *Negotiated Experiences* is fostering a movement inward for the *Mutuality of Engagement* dimension. There are some outward interactions of *Mutuality of Engagement-Negotiated Experiences*, mostly for PISEC and Science Theatre which are related to instances where facilitators were having a hard time engaging the children with the activities.

However, the facilitators expressed that positive interactions would always outweigh the inevitable frustrations.

We observe that *Mutuality of Engagement* is the least coded *Community Dimension* for Quavers to Quadratics and Science Theatre. One reason for this smaller amount of codes is related to identifying who is considered a member of the community of practice. In the case of Science Theatre and Quavers to Quadratics, the children are not regarded as members of the community, while for PISEC they are, which comes from the design of the programs. In PISEC, facilitators interact with the same group of children for one hour a week, for at least one semester, developing important bonds and relationships. The children are co-creators of the activities because they decide what they want to do and how, and the facilitator is there to provide support. The format of Science Theatre and Quavers to Quadratics means that children interact with the facilitators for a very short period of time and while there is some degree of participation from the children in how the activities developed during that time, there is more structure established by the facilitators on what will be done and how.

However, both in Science Theatre and Quavers to Quadratics facilitators spend more time working together as they develop and practice the activities or during the trips. This is evident from the main themes found in the *Mutuality of Engagement* codes for Science Theatre and Quavers to Quadratics. The themes referred to instances in which facilitators interacted with other facilitators, because they felt supported by their peers when learning the norms, content, and practices of the program. In the case of Science Theatre, newer facilitators expressed a general sense of support by the officers and returners because all of the more experienced members were happy to provide guidance and help where needed. For example, when Daisy is asked what has participating in Science Theatre meant to her, she talks about her interactions with other members:

> I went with [Jean] -- but she has been to other UP trips. She's super enthusiastic about all of the demos that we do, and it was really too cool to have her as my first person to have a show with, 'cause she taught me a lot [sic].

This was coded as *inbound Mutuality of Engagement* and *inbound Learning Trajectory* because for Daisy it is clear that Jean's enthusiasm for the program and the mentorship that she provided helped Daisy feel supported and build her membership in the community. Similarly, in the case of Quavers to Quadratics, the majority of the codes (93%) were associated with an inward trajectory within the community related to the facilitators bonding experience (see table in Figure 2), particularly throughout the Galway trip, as Orla expressed it:

> And it just get so and it's really nice to see and then in terms of like I guess the social side of things so it was really nice to bond as a group and I think we got really close and with the Galway people. It just worked really well, yes it was good [sic].

This was coded *insider Mutuality of Engagement* as it shows that Orla developed relationships with her fellow facilitators and enjoyed the bonding process, which means she views herself as a central member of the community. This experience had a positive effect on Orla's identity as shown by the *inbound Negotiated Experiences* code.

For both Science Theatre and Quavers to Quadratics, the trips (to rural Michigan and rural Ireland respectively) provided good opportunities for bonding outside of the regular practices of presenting the demos and facilitators mentioned how these interactions were positive in helping create connections inside and outside the community of practice. For example, Grace explains that bonds built between Science Theatre colleagues due to the time spent in the cars helped them create a rapport and trust that they relied on during the presentations causing codes of *Mutuality of Engagement inbound* and *Negotiated Experiences inbound*:

> Because we'd spend a lot of time in the car [laughter]. It's like no way you can't bond when you're in the car that much and spending that much time with each other. And during shows, things happen, so [laughter]. So if something goes wrong, you have to rely on each other [sic].

On the other hand, for PISEC the most relevant interactions come from the personal relationships that facilitators were able to build with the children over a semester(s), with the majority (65%) supporting an inward trajectory in membership. Lily discusses how interacting with the children and their enthusiasm makes her move to a more central position in the PISEC community but also influences her membership in the physics community:

> And then also just like being with the students in PISEC and like being around their excitement and enthusiasm helps me be, like I said earlier, kind of like more excited about physics. Like wow, this is really cool, like this is why I love physics is because of these, like, cool things that I get to learn with like middle school students [sic].

These thoughts from Lily were coded as *inbound Mutuality of Engagement* and *inbound Negotiated Experiences*. She then continues to say what she has gained from her participation, "I guess like as far as PISEC goes, just feeling like I'm a member of this community, like the CU middle and elementary schools. This kind of like community connection, just feeling like I'm a part of that is a big thing that I gained I would say." which was coded as *insider Mutuality of Engagement.*

Furthermore, facilitators described how the interactions with the children came as a refreshing change of scenery and perspective, compared to their everyday physics community of practice.

For example, Ava mentions how much she enjoyed having to work with a group of girls in PISEC because in her everyday physics PhD work she has mostly has male colleagues "[I]t was great because my group was always girls, and I really loved that because like here I'm just interacting with males, males, males all the time." This was coded as *Mutuality of Engagement inbound* and *Negotiated Experiences inbound* as Ava's positive interactions with members of the PISEC community moved her to a more central member of the PISEC community of practice and strengthened her physics identity.

Unlike in Science Theatre and Quavers to Quadratics, interactions with fellow facilitators were not discussed much in PISEC. In PISEC, facilitators do not have to work together, as each facilitator leads their own group. Facilitators only share car journeys to and from schools and the occasional social event, thus they seem to not affect each other's membership in the community. However, other interactions that did affect PISEC facilitators' membership were with their PhD advisors, in both inward and outward directions. For example, Evan describes his movement into the PISEC community, highlighted by *inbound Mutuality of Engagement* and *inbound Negotiated Experiences* codes, was originally prompted by his advisor "so it started out with my boss telling me that I should volunteer for PISEC, but as it went on I really started to kind of enjoy the interaction with the students." This is important because facilitators already struggle with conflict between PhD and PISEC commitments and are affected by the level of support their advisors give to their participation in PISEC. Encouragement from their advisors indicates to the facilitators that public engagement is part of the practices of the physics community of practice. Therefore, supervisors' support can impact the students' physics identity and their participation in the physics community. Unfortunately, some PISEC facilitators did not receive this support, like Ava who said she "never really had a lot of positive- in fact my advisor was kind of like not that supportive of doing outreach, because he was like 'yeah it's good but you should get your PhD first' kind of thing [sic]" which was coded as *outbound Mutuality of Engagement* and *neutral Nexus of Multimembership*. However, Ava continued with PISEC as "it was really fun. And the kids were like really enthusiastic, so that was great" [sic] which was coded with *inbound Mutuality of Engagement* and *inbound Negotiated Experiences.* Ava's experience shows how lack of advisor support did make her doubt her participation in PISEC but because she enjoyed working with the enthusiastic children she continued in the program, which resulted in her feeling a greater sense of belonging both in the PISEC and physics communities. PISEC facilitators' interactions with the children generally had an impact on their sense of belonging within both the PISEC and physics communities. Even though *Mutuality of Engagement* codes represented only 25% of the total *Community Dimensions*, those codes were meaningful within the PISEC community and more broadly with their membership within the physics community.

## Interactions with *Negotiability of the Repertoire*

*Negotiability of the Repertoire* is associated with members' competencies and knowledge in the skills, practices, and norms of the community. In Figure 2 we observe that *Negotiability of the*

*Repertoire* is the most represented *Community Dimension* for Science Theatre (42%) and Quavers to Quadratics (46%). In contrast, this mechanism is the least represented *Community Dimension* for PISEC (22%). This indicates that facilitators in Science Theatre and Quavers to Quadratics measure their membership mainly through their knowledge and competence of the repertoire as explained by Eoin (a third-year physics undergraduate student) from Quavers to Quadratics when asked whether he identifies as a musician. He starts by saying that he has performed with an orchestra and that he was able to do that because he had the set of skills necessary to participate in that community "[Y]our familiarity with your instrument and other music and how you apply it to new music sort of makes you a musician. And so I'd say that applies to me [sic]" then he goes on to say "Yeah I'd say so, and there's a sort of a general familiarity that you get when you spend enough hours doing something. And, like, I couldn't tell you how many hours I've put it into music [sic]." He also reflects that the same applies to his identity as a physicist "it's sort of parallels to the physicists [sic]." This was coded as *insider Negotiability of the Repertoire* because he is clearly describing his positionality in the community of practice through his knowledge and understanding of the practices of the community.

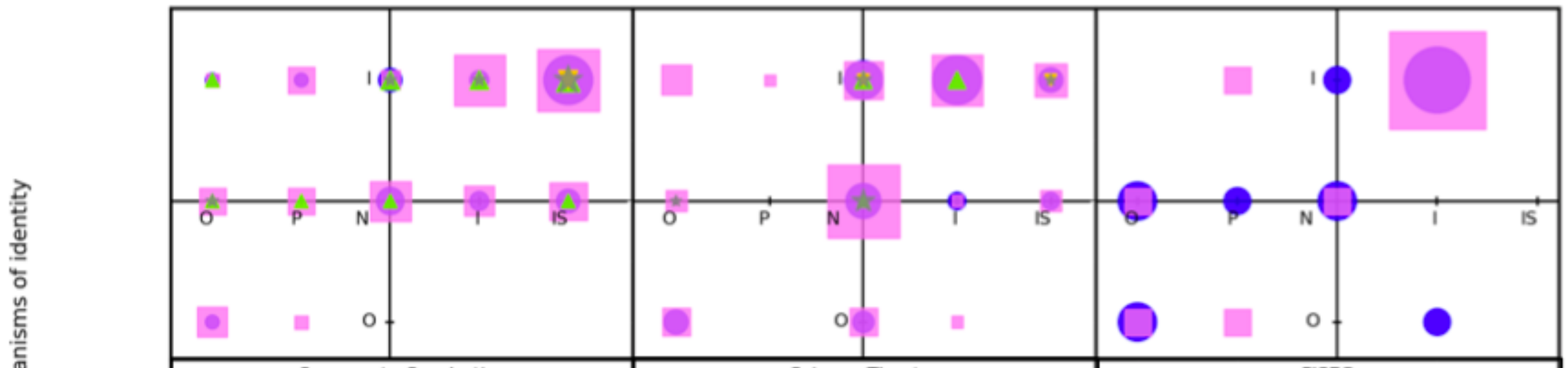

| Mechanisms of identity | | Quavers to Quadratics | | | | | Science Theatre | | | | | PISEC | | | | |
|---|---|---|---|---|---|---|---|---|---|---|---|---|---|---|---|---|
| | | Outbound | Peripheral | Neutral | Inbound | Insider | Outbound | Peripheral | Neutral | Inbound | Insider | Outbound | Peripheral | Neutral | Inbound | Insider |
| Inbound | NE o | 0.8 | 0.8 | 2.3 | 1.5 | 10.0 | 0.0 | 0.0 | 6.3 | 10.1 | 2.5 | 0.0 | 0.0 | 3.0 | 18.2 | 0.0 |
| | CM □ | 0.8 | 3.1 | 1.5 | 10.8 | 16.2 | 3.8 | 0.6 | 6.3 | 10.8 | 4.4 | 0.0 | 3.0 | 0.0 | 39.4 | 0.0 |
| | LT △ | 0.8 | 0.0 | 1.5 | 1.5 | 2.3 | 0.0 | 0.0 | 1.3 | 1.3 | 0.0 | 0.0 | 0.0 | 0.0 | 0.0 | 0.0 |
| | RbLG ▽ | 0.0 | 0.0 | 0.0 | 0.0 | 1.5 | 0.0 | 0.0 | 0.6 | 0.0 | 0.6 | 0.0 | 0.0 | 0.0 | 0.0 | 0.0 |
| | NM ☆ | 0.0 | 0.0 | 0.8 | 0.8 | 3.8 | 0.0 | 0.0 | 1.3 | 0.0 | 0.6 | 0.0 | 0.0 | 0.0 | 0.0 | 0.0 |
| Neutral | NE o | 0.0 | 0.0 | 3.1 | 1.5 | 2.3 | 0.0 | 0.0 | 5.1 | 1.3 | 1.3 | 6.1 | 3.0 | 6.1 | 0.0 | 0.0 |
| | CM □ | 3.1 | 3.1 | 6.9 | 3.8 | 6.2 | 1.9 | 0.0 | 22.2 | 0.6 | 1.9 | 3.0 | 0.0 | 3.0 | 0.0 | 0.0 |
| | LT △ | 0.8 | 0.8 | 0.8 | 0.0 | 0.8 | 0.0 | 0.0 | 0.6 | 0.0 | 0.0 | 0.0 | 0.0 | 0.0 | 0.0 | 0.0 |
| | RbLG ▽ | 0.0 | 0.0 | 0.0 | 0.0 | 0.0 | 0.0 | 0.0 | 0.0 | 0.0 | 0.0 | 0.0 | 0.0 | 0.0 | 0.0 | 0.0 |
| | NM ☆ | 0.8 | 0.0 | 0.0 | 0.0 | 0.0 | 0.6 | 0.0 | 2.5 | 0.0 | 0.0 | 0.0 | 0.0 | 0.0 | 0.0 | 0.0 |
| Outbound | NE o | 0.8 | 0.0 | 0.0 | 0.0 | 0.0 | 2.5 | 0.0 | 1.9 | 0.0 | 0.0 | 6.1 | 0.0 | 0.0 | 3.0 | 0.0 |
| | CM □ | 3.8 | 0.8 | 0.0 | 0.0 | 0.0 | 3.2 | 0.0 | 3.2 | 0.6 | 0.0 | 3.0 | 3.0 | 0.0 | 0.0 | 0.0 |
| | LT △ | 0.0 | 0.0 | 0.0 | 0.0 | 0.0 | 0.0 | 0.0 | 0.0 | 0.0 | 0.0 | 0.0 | 0.0 | 0.0 | 0.0 | 0.0 |
| | RbLG ▽ | 0.0 | 0.0 | 0.0 | 0.0 | 0.0 | 0.0 | 0.0 | 0.0 | 0.0 | 0.0 | 0.0 | 0.0 | 0.0 | 0.0 | 0.0 |
| | NM ☆ | 0.0 | 0.0 | 0.0 | 0.0 | 0.0 | 0.0 | 0.0 | 0.0 | 0.0 | 0.0 | 0.0 | 0.0 | 0.0 | 0.0 | 0.0 |

Negotiability of the Repertoire

**Figure 5:** The above figure shows how the *Mechanisms of Identity* overlap with the *Negotiability of the Repertoire*. The frequency of the overlapping is represented by the size of the markers. The table mirrors the plot and shows the frequencies in each quadrant marked with different colors.

***Community Membership*** is shown in Figure 5 as having the most interactions with *Negotiability of the Repertoire (*shown by the purple squares). *Community Membership* relates to the learning of the community's practices through interactions with other members or experiences that happened while engaging in the practices of the community. As *Negotiability of the Repertoire* is the most frequently coded *Community Dimension* for Science Theatre and Quavers to Quadratics,

*Community Membership* is also the most frequently coded *Mechanism of Identity* for those two programs (39% and 34% respectively).

Both Science Theatre and Quavers to Quadratics are designed to place emphasis on learning the material and how to do the activities for the university facilitators to become more central members of the community. For example, both of those programs have challenging activities requiring the facilitators to take up more central roles in order to demonstrate activities and explain the phenomena behind those, therefore putting more perceived emphasis on content knowledge. There is also an emphasis put on the performative aspect, which is mentioned by Liam in Science Theatre:

> There [were] a lot of demonstrations where I kind of understood them. Like, when someone explained it to me, I'm like yeah, that makes sense. But it's a whole different thing, someone explaining this is how you derive X as opposed to you explaining this is how you derive X [sic].

This reflection from Liam was coded as *inbound Negotiability of the Repertoire* because Liam is explaining how he felt that there is a difference in understanding the content and explaining the content, especially if it is to younger kids. Therefore, he is moving towards becoming a central member as he develops those skills. Liam's identity within Science Theatre is growing as his competence with the practices develops, captured by the *inbound Community Membership* code.

While there is considerable spread of interactions, the largest concentration occurs in the upper-right quadrants of Figure 5 (*inbound* and *insider* codes). This indicates that participation in the practices of the community and the process of the learning of the practices through interactions with other members supports and fosters facilitators' identity. Ciara from Quavers to Quadratics shares:

> I'm getting a lot of confidence. I like personal confidence and the fact that you're not like... you don't shy away from the fact that you're starting off. I didn't know I was you know sound waves or frequency [are] or anything like that. You had to just take the challenge and it's really rewarding to see the kids understand different concepts [sic].

This was coded as *inbound Negotiability of the Repertoire* and *inbound Community Membership* as Ciara's positive approach to learning the physics content so she could successfully teach it to the children caused an increase in her personal confidence. This confidence boost produced an inward movement in her membership in both the Quavers to Quadratics community and the physics community.

Science Theatre members build and interact with their repertoire by often speaking about the demos and activities unique to the program and how they have discovered better ways to communicate. For example, Jacob is talking about how using certain expressions or talking about specific concepts do not necessarily add to the understanding of the phenomena:

> [W]e have another one, another Physics demo, that you just don't bring up mass, because mass is-- you can say in all these equations, there is mass, but trying to get a little kid to understand what mass is, is not worth it. And all the mass cancels out anyway [laughter], so you're never going to be, 'Okay, we've done this experiment, having Sally spin around on the chair. Now we're going to do a massless spinning.' Like no, that's-- so yeah, it's just figuring out what to bring up and stuff like that. And so I think as the week went on, I definitely got a better sense of like, okay, judging my audience and figuring out what they could handle [sic].

This was coded as *inbound Negotiability of the Repertoire* because he is describing how through the process of participating and engaging in the practices of the community, he was developing more confidence and competence in those practices. Jacob's growing confidence affects his identity positively hence it was also coded as *inbound Community Membership***.**

*Negotiability of the Repertoire* represents only 22% of the *Community Dimensions* in PISEC (see Figure 2 for details). Unlike in Science Theatre and Quavers to Quadratics, the PISEC facilitators are graduate physics students. As practitioners we know that knowledge of physics content is less important to graduate student facilitators compared to program values because they already feel confident enough about the physics practices of the community. The physics content is part of their repertoire in the physics community of practice, making it easier for the facilitators to incorporate physics content from their membership in the physics community of practice into the PISEC community of practice. Also, the model of the PISEC program requires less performance than the other programs, where the facilitator would be in the role of speaking in front of a passive audience. Rather, PISEC activities are more exploratory and one-on-one, with facilitators working alongside children on hands-on activities to try to answer questions. Thus, in PISEC, facilitators do not need to act as experts in front of a crowd but rather as coaches with the children, which additionally de-emphasizes the importance of expert physics content knowledge. This is reflected in the absence of codes related to learning the content- while it was the case for both Science Theatre and Quavers to Quadratics facilitators.

PISEC facilitators focus on the practices related to learning to communicate the content appropriately for their audience, as discussed by Lily:

> Well one thing, I think it's really important for physicists to be able to explain their work, and physics concepts in general, in a simple way to young kids or to people

> who are not familiar with physics, and that's really hard to do. So like when you first do PISEC and you have to do the videos- I think like it just takes a bunch of practice I think. And I often find like family members or friends who never learned physics will ask me stuff about physics, like 'what is physics, what do physicists do, what do you do?' and sometimes it's really hard for me to explain it in a way that is accessible. So I think being involved in these types of programs is really helpful for that [sic].

Lily's comments showed a growing understanding of the PISEC practices and a developing confidence about how she can explain physics, so this was coded as *inbound Negotiability of the Repertoire* and *inbound Community Membership.* Knowing how to communicate with the audience is seen as the most important practice for being a member of the PISEC community, but also identifying those practices as part of the physics community repertoire. This value system of skills naturally affects participation in both communities of practice, PISEC and the physics community. Therefore, the improvement in those practices produces inward movement into those communities.

We should note, *Negotiability of the Repertoire* has the largest percentage of outbound codes for all three programs - representing less than 20% of the overall *Community Membership* codes (see table in Figure 2 for details). The outbound codes mainly referred to the challenge of developing practices that lie outside a facilitator's area of expertise. In Quavers to Quadratics, Ciara is a facilitator with no science background and expressed "So we were actually a team of musicians put together but at first it was a bit daunting, with the science aspect [sic]" which was coded as *Negotiability of the Repertoire outbound* and *Community Membership inbound*. However, she then she continues to say "actually it worked to our advantage because we found our own way and we kind of made our lesson plan to see the way that we were looking at things [sic]" which was coded *Negotiability of the Repertoire insider* and *Community Membership inbound* showing the development of Ciara's confidence with the Quavers to Quadratics practices. In Science Theatre, the large majority of outbound codes came from a student without a science background, so it is expected that he might see the learning of the practices harder than someone with already some knowledge of the content. In general, the outbound codes tended to come from newer members of the community, demonstrating the need to support newcomers and scaffold their inward trajectory within the community of practice.

In summary, the interaction between *Negotiability of the Repertoire* with *Community Membership* is perceived as most relevant to membership for Science Theatre and Quavers to Quadratics facilitators, possibly as a result of the program designs. In these programs, the facilitators demonstrate the different physics phenomena that the children are observing, so there is more specific content knowledge for them to learn. Additionally, in both cases the facilitators are undergraduate students coming from different backgrounds - not just physics or even science - needing to learn the content to reach the point of confidence to deliver the activities. Hence, *Community Membership* seems to be an important mechanism for building identity. This is in agreement to previous studies on physics identity [7] which indicate that competence/performance is one of the three main determinants of a person's discipline-based identity, i.e., the more competent an individual feels in the practices, the more that individual identifies themselves as a member of the community.

# DISCUSSION

Our analysis of the three presented informal physics programs has shown that there are particular structures and practices that support physics identity growth. In Figure 7, we present the main themes from our analysis of facilitators' identity development due to their participation in the programs. These five themes or program structures were seen throughout the interview data. Additionally, the operationalized Community of Practice framework highlights the potential challenges facilitators experience in the program and can prompt practitioners to enact structural changes within their own programs. The specifics of the necessary changes to a program structure will naturally be different for each program due to the differing practices and community needs. Therefore, the details of the specific suggested changes to the three analyzed programs are outside the remit of this paper, but we will discuss how the themes of the programs highlighted by the coding framework might generally change the programs' structure.

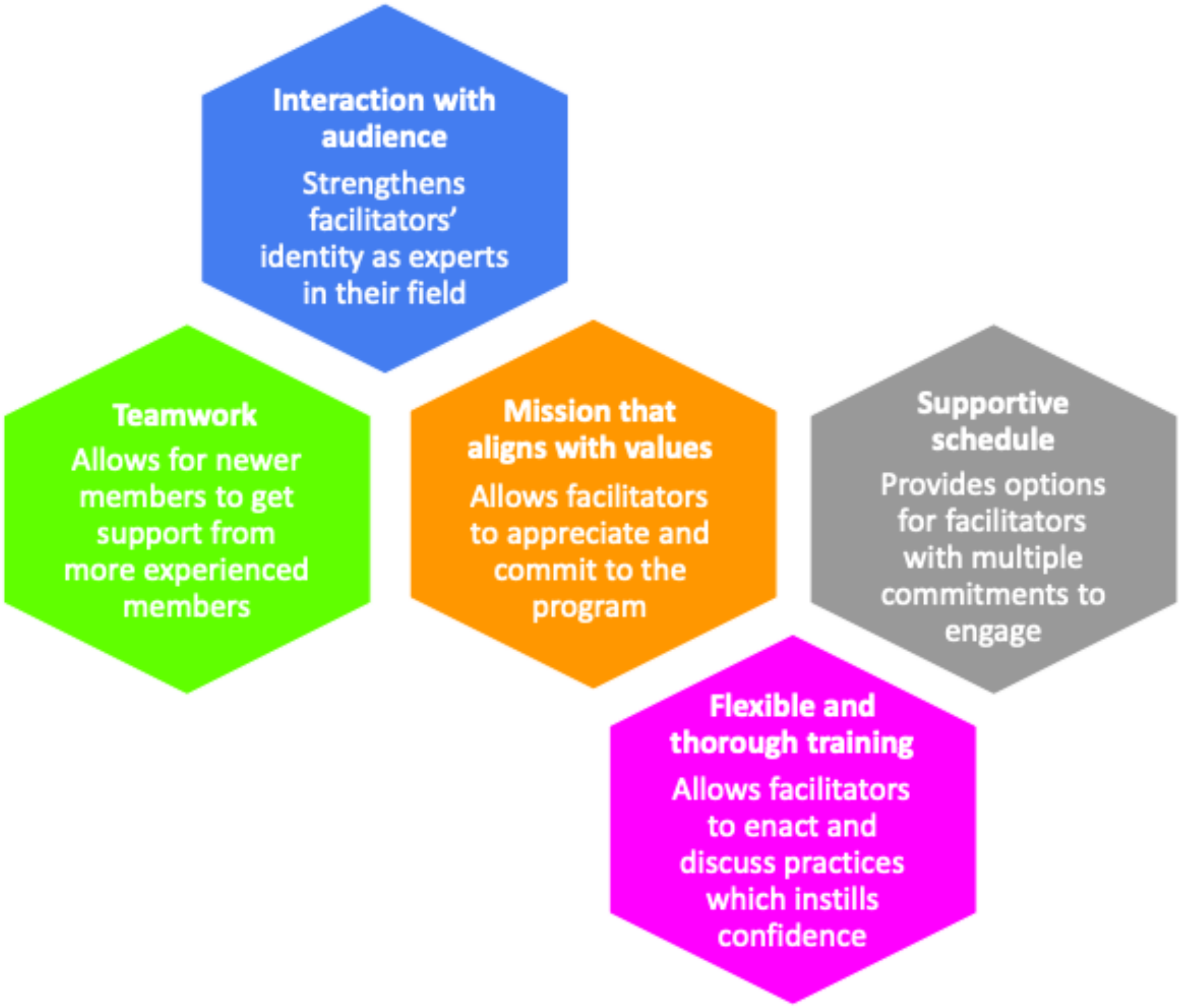

**Figure 7:** Identified themes that support facilitators' identity based upon the interview analysis. These structures are general, and practitioners are encouraged to consider other program specific features. The color of each hexagon indicates the *Community Dimension* or *Mechanism of Identity* that each structure is connected to (e.g. Teamwork is in green as it connects to *Mutuality of Engagement*).

A major theme across the three programs highlighted by the *Accountability to the Enterprise* code and the *Relationship between Local and Global* code is the facilitator's desire to participate in informal science programs. These programs are a way for physics students to have a physics experience outside of their classes and labs that aligns with their own values. Many facilitators discussed their desire to participate in informal programs prior to joining their current program, some even selecting the university they attended based on the informal programs available. As a whole, the facilitators were highly committed to their respective programs as they valued the teaching and acting as role model opportunities that came with participation in each program. Facilitators expressed a desire to share their physics knowledge and their passion for physics with younger students, especially those who are underrepresented or who may not have access to physics due to rural settings or socioeconomic status. Furthermore, many facilitators spoke about how they did or did not have access to informal programs when they were young and how this inspired them to make sure others did have this opportunity.

Within the interviews, the benefits from participation identified by the facilitators consist of: productive interactions with the children, learning communication skills, gaining additional content knowledge, and being part of a community of people with similar goals. We can further expand this list to include a more supported passion for physics due to the children's positive reactions, a sense of contribution to the field when the children express the desire to continue with science, a greater sense of confidence as they are recognized as experts in physics, a greater sense of belonging in the community, and a widening of their perception of who and what belongs in physics. Across all three programs, the facilitators expressed a sense of enjoyment and a desire to participate again, particularly because participation in these programs connected to their discipline and supported their membership in the physics community of practice. This evidence demonstrates how informal physics programs are valued by facilitators and should act as a prompt to continue and expand informal physics programs similar to the programs analyzed in this paper. Current and future informal practitioners should be encouraged by the fact that we see facilitators recognizing the benefits of participating in informal practices and are highly committed to the values of informal physics experiences.

Another identified theme involved direct engagement with the children that prompted positive identity development for the facilitators. We saw this theme within the *Negotiated Experiences* identity mechanism and all three *Community Dimension* codes, particularly *Accountability to the Enterprise.* The facilitators connect with the missions of these programs, which broadly is showing that physics is fun and is for anyone, because of their membership in the physics community and established discipline-based identity. They commit because the missions align with their identities. Having positive reactions to their enactment of these missions moves them inwards in the informal communities and the physics communities while strengthening their physics identities because of this connection. The facilitators really enjoyed building relationships with the kids and seeing the effect that their participation in outreach has for the children. We suggest that any developments in the programs' structures should continue to place the interactions between the children and facilitators as the most important aspect of the program. The framework shows that helping others, in this case children, build a physics identity can strengthen facilitators' identity, similar to what has been shown in the formal setting with Learning Assistants [29-30]. Therefore, programs like Quavers to Quadratics, Science Theatre, and PISEC should continue to be created and sustained not only for the benefit they have for the children involved in the outreach but also for the facilitators providing the outreach.

One identified challenge considers how the facilitators worry about fitting outreach into busy schedules. We saw this theme through the *Nexus of Multimembership* code and a few *Accountability to the Enterprise* codes as some facilitators experienced conflict between their desire to participate in the program and their busy academic schedules. They explain how this time spent away from academics is always beneficial and they wish to commit fully to the programs but are worried about not having enough time for school. This worry about not having the time to take part in the programs caused a negative impact on the facilitators' identities. We recommend that greater support from both the programs and their academic supports (i.e. physics departments, research advisors, professors, etc.) would be appreciated by student facilitators. Practitioners of informal programs can help with this issue by being aware of their facilitators' other commitments and trying as much as possible to work around them. Our analysis showed the facilitators really wanted to commit to informal programs as they saw the benefits, but time was a barrier to participation.

Another challenge that came up often in the qualitative analysis of the interviews, and was validated by the coding, was the anxiety facilitators initially felt through the learning and implementation of the practices in the programs. This was mainly due to the performative nature of the programs' practices and a requirement to learn content knowledge adjacent to their background degree, particularly in Science Theatre and Quavers to Quadratics. The coding framework showed us (through the frequency and the subcodes of the *Negotiability of the Repertoire* and *Community Membership* codes) how having a strong working knowledge of the practices of the communities was important to facilitators and the anxiety they felt about their perceived lack of knowledge did negatively affect identity development and their position within the community. We posit that a change to the structure of the programs' education system is required to instill greater confidence in the facilitators. We believe educational sessions that allow all facilitators the opportunity to enact and discuss the practices of the program is a key adjustment necessary to aid facilitators in increasing their confidence with content. Program education should be carefully considered and consistently adjusted to facilitators' needs especially when practices change or expand. We would like to note that facilitator education is something that continually evolves, and the three programs have since altered or improved education strategies after this data was collected. The education in place during this data collection aimed to instill confidence in the facilitators but our analysis shows that all three programs had room for improvement within their facilitator education. Some changes that have taken place since this research was conducted include PISEC creating facilitator training that emphasizes the pedagogical component and Quavers to Quadratics creating training days for the facilitators before they go into a classroom.

Teamwork is a structure already in place in Quavers to Quadratics and Science Theatre that helps ease anxiety and support the facilitators. We see through the *Mutuality of Engagement* code and *Negotiated Experiences* code how new facilitators appreciate working with more experienced community members as they can support them in their learning of the program practices. As well as increasing their confidence with the practices of the program, working with more experienced facilitators also moved the new facilitators inwards in the community. The bonding between facilitators that took place on both Science Theatre's and Quavers to Quadratics' trips caused positive identity development and further shows how the interaction between members of these groups is important. Teamwork is an element of the programs' structures that has positive impacts on both the facilitators' position in the community of practice and on their identity development. Therefore, we recommend that there should be a focus to embed teamwork further into the structure of informal physics programs in order to create a stronger, close-knit community.

In summary, using the Communities of Practice framework we have found that three different informal physics programs provided spaces for positive physics identity development among the student facilitators. Participating in informal programs supported the facilitators' interest in physics, allowed them to be recognized as members of the physics community, and expanded and supported their participation in the practices of the physics community. The coding identified challenges faced by the facilitators and how some program structures could be improved such as adjusting education to increase facilitators' confidence with the practices and ensuring facilitators' multiple time commitments are considered. The framework highlighted the desire facilitators had to participate in science outreach, their appreciation for working with more experienced facilitators, and their enjoyment of working with children. The framework also showed how informal programs can support the development of a more inclusive and diverse physics community of practice by supporting facilitators' multiple identities, including their physics identity.

# IMPLICATIONS AND CONCLUSIONS

In this paper, we have shown how structures of informal physics programs can be identified using the Communities of Practice framework in order to understand discipline-based identity development of student facilitators. As the framework simply requires a program to operate as a community of practice – a group of people working together (a community) to achieve a shared mission (a domain) through various activities (practices) - it can be used by practitioners to analyze appropriate programs. We have shown that the operationalized framework is a valuable tool, able to inform researchers and practitioners of the structures of informal programs that support facilitators and how their involvement affects identity development. The differences and similarities between these three programs were highlighted by the framework coding results, as discussed in the analysis, showing that the framework can adapt to complexities seen in various programs.

We envision this operationalized framework being useful for many contexts outside of informal physics programs and provide an example of how one might use it to analyze other aspects of physics identity development. There are many communities of practice in action in the daily life of a physicist. For example, the physics department of a university has a domain that includes furthering physics research and educating the physicists of the future, a community of physicists, and a practice involving giving lectures and running labs for students, conducting experiments and other research, participating in seminars and conferences etc. This framework could be used to analyse how central the members of that community of practice feel within the physics department community and how they are (or are not) supported with the continuous development of their identity as a physicist. This in turn would show how the physics department might need to adjust structures and practices to ensure all members feel central in the community and experience positive identity development.

In closing, we have shown how structural elements of three different informal physics programs, after undergoing analysis by the operationalized Communities of Practice framework, prompted positive physics identity development among the student facilitators. We have provided some recommendations for ongoing and future informal physics programs. Finally, we believe that the identified themes and the operationalized framework are applicable to many potential research projects and communities of practice that impact physics identity.

# ACKNOWLEDGEMENTS

We would first like to acknowledge and thank our study participants, who have shared their time, energy, and stories to contribute to this work, as well as all three programs (PISEC, Quavers to Quadratics, and Science Theatre) for being open to participating in this work. Additionally, we want to acknowledge Manuel Vazquez for his support on the initial stages of this study and Noah Finkelstein for his guidance and contribution to all of this work. This paper is based upon work supported by the European Research Council through Marie Skłodowska-Curie Actions Individual Fellowships (MSCA-IF) project no. 794434, and NSF Advancing Informal STEM Learning award #1423496. Any opinions, findings, and conclusions or recommendations expressed in this material are those of the author(s) and do not necessarily reflect the views of the European Research Council or the National Science Foundation.